\documentclass[letterpaper,prb,reprint,superscriptaddress,floatfix]{revtex4-2}
\usepackage{lineno}
\usepackage{graphicx}
\usepackage[dvipsnames]{xcolor}
\usepackage{graphics}
\usepackage{longtable}
\usepackage{url}
\usepackage{bm}
\usepackage{amsmath}
\usepackage{titlesec}
\titlespacing\section{0pt}{18pt plus 4pt minus 2pt}{18pt plus 2pt minus 2pt}
\usepackage{footmisc}
\usepackage[colorlinks=true, citecolor=blue, linkcolor=blue]{hyperref}

\usepackage[normalem]{ulem}

\begin{document}

\title{Evidence for ferroaxial order in 1T-TiSe$_2$ via elastoresistivity measurements}

\author{Qianni Jiang}
\affiliation{Department of Applied Physics and Geballe Laboratory for Advanced Materials, Stanford University, Stanford, CA 94305}
\affiliation{Stanford Institute for Materials and Energy Sciences, SLAC, Menlo Park, CA 94025}
\author{Ezra Day-Roberts}
\affiliation{Department of Physics, Arizona State University, Tempe, AZ 85287, United States}
\affiliation{Department of Chemical Engineering and Materials Science, University of Minnesota, Minneapolis, Minnesota 55455, United States}
\author{Benito Gonzalez}
\affiliation{Department of Applied Physics and Geballe Laboratory for Advanced Materials, Stanford University, Stanford, CA 94305}
\affiliation{Stanford Institute for Materials and Energy Sciences, SLAC, Menlo Park, CA 94025}
\author{Awadhesh Das}
\affiliation{Department of Physics, Temple University, Philadelphia, Pennsylvania 19122, USA}
\author{Darius H. Torchinsky}
\affiliation{Department of Physics, Temple University, Philadelphia, Pennsylvania 19122, USA}
\author{Turan Birol}
\affiliation{Department of Chemical Engineering and Materials Science, University of Minnesota, Minneapolis, Minnesota 55455, United States}
\author{Rafael M. Fernandes}
\affiliation{Department of Physics, The Grainger College of Engineering, University of Illinois Urbana-Champaign, Urbana, Illinois 61801, United States}
\affiliation{Anthony J. Leggett Institute for Condensed Matter Theory, The Grainger College of Engineering, University of Illinois Urbana-Champaign, Urbana, Illinois 61801, United States}
\author{Ian R. Fisher}
\email[Corresponding author email: ]{irfisher@stanford.edu}
\affiliation{Department of Applied Physics and Geballe Laboratory for Advanced Materials, Stanford University, Stanford, CA 94305}
\affiliation{Stanford Institute for Materials and Energy Sciences, SLAC, Menlo Park, CA 94025}

\date{\today}
\begin{abstract}
         The study of spontaneous symmetry breaking and electronic order is fundamental in condensed matter physics. Hidden order, that is, symmetry-breaking states that elude conventional probes, potentially plays a crucial role in understanding complex quantum phases in a wide range of materials. Ferroaxial order, a state characterized by broken mirror symmetries while maintaining time-reversal and inversion symmetries, is one of the hidden orders that have proven most challenging to detect experimentally. In this work, we demonstrate a new approach for investigating both the ferroaxial order parameter and the associated ferroaxial susceptibility using elastoresistivity measurements. We do this for the transition metal dichalcogenide 1T-TiSe$_{2}$, a material that exhibits charge density wave order that has eluded comprehensive understanding for a long time. These measurements reveal an anomalous antisymmetric off-diagonal linear elastoresistivity in the CDW state. We discuss why this provides a smoking gun for ferroaxial order. Furthermore, we construct an appropriate cubic combination of the symmetry-breaking strains $\epsilon_{x^2-y^2}$ and $\epsilon_{xy}$ that acts as an effective conjugate field for the ferroaxial order, and demonstrate how sweeping this effective field back and forth in the CDW state results in a hysteretic behavior of the elastoresistivity, which we conclude is associated with the movement of ferroaxial domain walls. Finally, we reveal a divergence of certain nonlinear elastoresistivity coefficients above the critical temperature, and discuss how this is consistent with a divergence of the ferroaxial susceptibility near the charge density wave transition temperature (T$_{\rm{CDW}}$ $\sim$ 200K). Our study also includes detailed elastocaloric effect measurements, which reveal the presence of an additional phase transition several tens of Kelvin below T$_{\rm{CDW}}$. We also use optical second harmonic generation to carefully exclude the possibility of chiral order. Our results provide important new insights about the symmetry of the ordered state in 1T-TiSe$_{2}$. This work also highlights the potential of elastoresistivity as a valuable tool for revealing the presence and symmetry of certain hidden order states.

\end{abstract} 
\pacs{}
\maketitle

\section{Introduction}

\begin{figure*}[t!]
\centering
\includegraphics[trim={0 0cm 0cm 0cm},clip,width=0.95\textwidth]{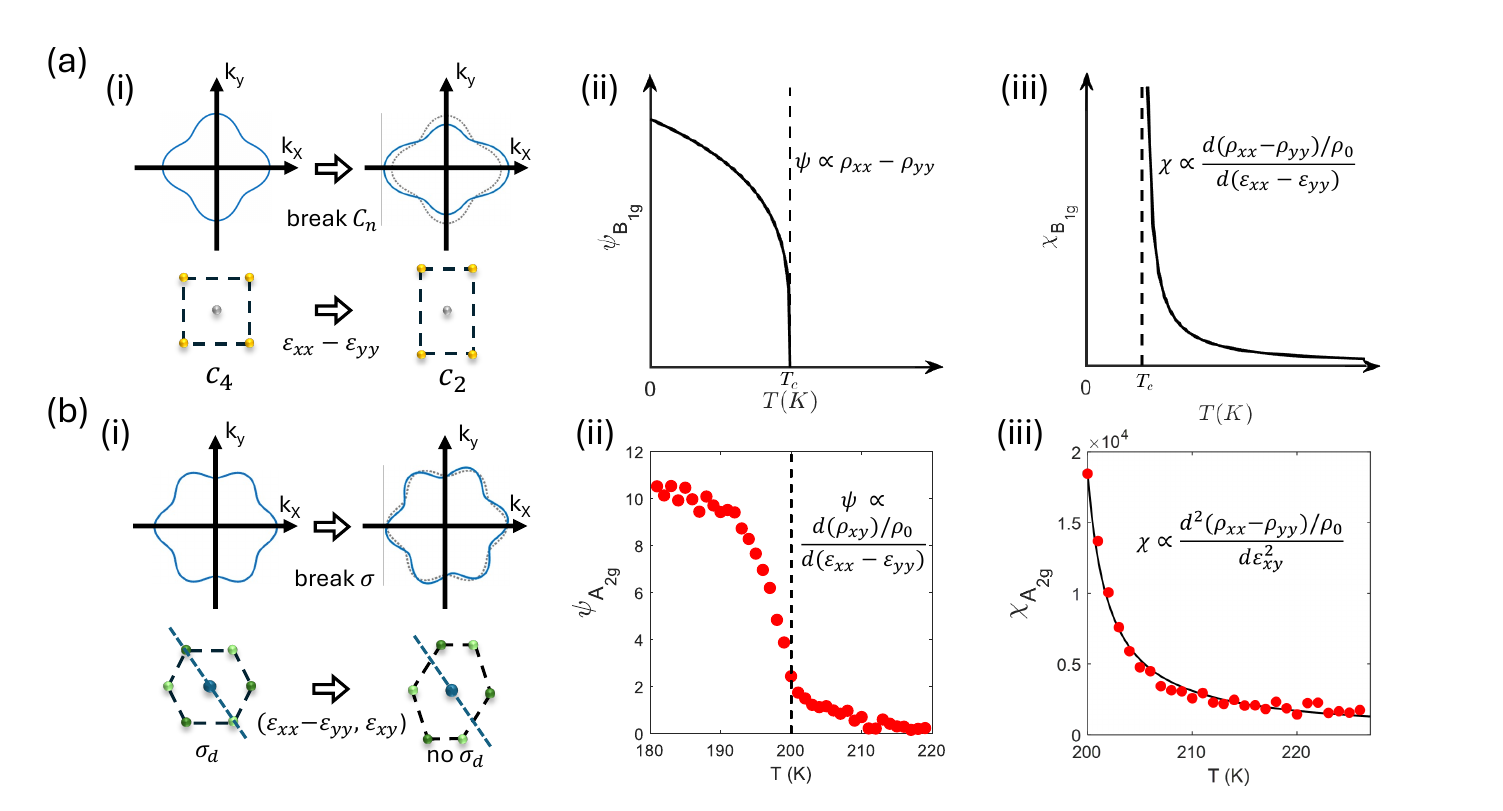}
\caption{\textbf{Illustrating how strain can act as an effective conjugate field to couple to (a) nematic and (b) ferroaxial order}. (a) The case of nematic order in a tetragonal lattice. (i) Schematics of a hypothetical Fermi surface in a tetragonal material for temperatures above (left) and below (right) the nematic transition, which breaks the C$_{4}$ rotational symmetry of the tetragonal lattice, here with an $x^2-y^2$ ($B_{1g}$) symmetry. Below, schematic illustration of in-plane anisotropic strain $\epsilon_{xx}-\epsilon_{yy}$, which has the same symmetry as the nematic order parameter, and is therefore an effective conjugate field for the nematic order parameter. (ii) Schematic showing the temperature-dependence of the nematic order parameter, $\psi_{B_{1g}}$, which is proportional to the resistivity anisotropy $\rho_{xx} -\rho_{yy}$ close to the critical temperature, $T_c$. (iii) Schematic showing the temperature-dependence of the nematic susceptibility $\chi_{B_{1g}}$, which can be probed by the linear elastoresistivity $\frac{d(\rho_{xx}-\rho_{yy})/\rho_{0}}{d(\epsilon_{xx}-\epsilon_{yy})}$ and which shows divergent behavior above the nematic transition. (b) The case of ferroaxial order in a trigonal lattice. (i) Schematics of a hypothetical Fermi surface for temperatures above (left) and below (right) the ferro-axial transition, which breaks the vertical mirror symmetry $\sigma_d$. Only a cubic combination of the deviatoric strains $\epsilon_{xy} \left[ 3(\epsilon_{xx}-\epsilon_{yy})^2-4\epsilon_{xy}^2\right]$ has the correct symmetry to serve as a conjugate effective field for the ferroaxial order. (ii) Data for TiSe$_2$ showing the elastoresitivity coefficient $\frac{d\rho_{xy}/\rho_{0}}{d\epsilon_{xx}-\epsilon_{yy}}$ as a function of temperature, which reveals the temperature-dependence of the ferroaxial order parameter $\psi_{A_{2g}}$. (iii) Data for TiSe$_2$ showing the nonlinear elastoresistivity $\frac{d^{2}(\rho_{xx}-\rho_{yy})/\rho_{0}}{d^{2}\epsilon_{xy}}$, which for temperatures above the ferroaxial transition reveals the temperature-dependence of the ferro-axial susceptibility $\chi_{A_{2g}}$.}
\label{fig:fig1}
\end{figure*}

Ferroaxial order is a specific type of ordered state characterized by a uniform ($q=0$) rotational structural/electronic distortion and, thus, by an axial vector \cite{Hlinka2016,Hayami2018,Cheong2018,Winkler2023,DayRoberts2025}. A phase transition into such a state breaks all vertical mirrors while preserving any horizontal mirrors, certain rotational symmetries, inversion, and time reversal symmetry. Examples of materials that undergo a ferroaxial phase transition are extremely rare. The principle examples, RbFe(MoO$_4$) and NiTiO$_3$, are both understood to be structurally driven \cite{ferroaxialRFMO,ferroaxialNiTiO3,Zeng2025}. Electronically driven ferroaxial order has been proposed to occur in TaS$_2$ and ErTe$_3$ \cite{ferroaxialTaS2,Ken2025}. Unlike more commonly studied orders, ferroaxial order has remained challenging to detect experimentally, in part due to its subtle nature, and in part because coupling to it requires higher-order conjugate fields. Indeed, nature does not provide a single uniform effective field that has the right symmetry to couple to ferroaxial order (justifying its identification as a ``hidden order"), motivating development of tools and techniques that use composite effective fields to couple to and control this elusive broken symmetry state and its fluctuations. This is the essence of the work described in this manuscript. \\
\indent Nonlinear strain offers a powerful method to probe and control ferroaxial order. The approach is analogous to that employed in probing electronic nematic (i.e., ferroquadrupolar) order using linear strain (Fig. \ref{fig:fig1} (a)). Nematic order is a uniform ($q=0$) state that breaks a discrete rotational symmetry of the crystal lattice, while preserving time reversal symmetry. Importantly, because strain is also a symmetric rank-2 tensor, anisotropic strain that breaks the same crystal symmetry can serve as a conjugate effective field for the nematic order \cite{NematicityBa122, NematicFebased}. Furthermore, because the resistivity anisotropy is also a rank-2 tensor, it can be used as a measure of the nematic order parameter (Fig. \ref{fig:fig1} (a)(ii)) \cite{ResistiveanisotropyBa122}. Above the transition, the nematic susceptibility can then be probed by applying the conjugate effective field (i.e. anisotropic strain with the same symmetry as the nematic order parameter), and the induced nematic order parameter can be measured by the resistivity anisotropy it creates (Fig. \ref{fig:fig1} (a)(iii)) \cite{NematicityBa122,NematicFebased}. Similarly, inside the ordered state (below the critical temperature), applying the same strain can detwin nematic domains, producing hysteresis in the response of the order parameter to a conjugate field \cite{hystersisinBa122}. Thus strain, and elastoresistivity, can manipulate the nematic order parameter inside the ordered state, and play witness to fluctuations of the order parameter above the critical temperature respectively. 
\\
\indent A similar approach can be extended to investigate ferroaxial order, as illustrated in Fig. \ref{fig:fig1} (b). Formally, ferroaxial order corresponds to an electric toroidal dipole moment, which is nothing but a three-component axial vector. However, in certain crystals, an electric toroidal moment mixes with electric hexadecapole moments (corresponding to a $g$-wave state), suggesting that a rank-four tensor could be used to probe the ferroaxial order parameter. Of particular relevance to the present work, symmetry analysis has shown that specific components of the elastoresistivity tensor $m_{jlkl}= \frac{\partial\rho_{ij}}{\partial\epsilon_{kl}}$ can be used to detect the onset/presence of ferroaxial order \cite{DayRoberts2025}, as indicated in Fig. \ref{fig:fig1}(b)(ii)). In particular, and specializing now to the specific point group to which 1T-TiSe$_2$ belongs, the symmetry character of the ferroaxial order parameter (which transforms as the $A_{2g}$ irrep of the point group) necessarily turns an $x^2-y^2$ symmetry strain (transforming as the first component $E_g^{(1)}$ of the two-dimensional $E_g$ irrep) into an $xy$ symmetry resistive response (transforming as the second component $E_g^{(2)}$), and vice versa. This anomalous orthogonal response to anisotropic  strain can be used as a smoking gun for ferroaxial order, as pointed out in Ref. \cite{DayRoberts2025}. Furthermore, to effectively couple to this order parameter, a higher-order conjugate field is necessary, too. In the case of TiSe$_{2}$, which belongs to the D$_{3d}$ point group, a cubic combination of the two components of the two dimensional E$_{g}$ irrep (i.e. $\epsilon_{x^2-y^2} = \epsilon_{xx}-\epsilon_{yy}$, and $\epsilon_{xy}$) can serve as such a conjugate field. This suggests that nonlinear elastoresistivity can be used to probe the ferroaxial susceptibility \emph{above} the transition  (Fig. \ref{fig:fig1}(b)(iii)) while,  \emph{below} the transition, it reveals a hysteresis loop reflecting the motion of ferroaxial domains (Fig. \ref{fig:fig3}).
        
\indent Recent studies of 1T-TiSe$_{2}$ have raised intriguing questions about additional hidden symmetry-breaking near its 2$\times$2$\times$2 charge density wave (CDW) transition at 200 K \cite{TiSe2ChiralCDWXray, TiSe2gyrotropic, TiSe2XrayTwotransitions, TiSe2ChiralCDWKim2024, TiSe2chiralCDWSTM,  TiSe2nonchiralSHG}. Despite extensive research, the microscopic mechanism driving the CDW -- whether dominated by excitonic interactions or electron-phonon coupling -- remains unresolved. In addition, recent investigations using techniques such as the circular photogalvanic effect (CPGE) \cite{TiSe2gyrotropic}, scanning tunneling microscopy (STM) \cite{TiSe2chiralCDWSTM}, X-ray scattering \cite{TiSe2ChiralCDWXray,TiSe2XrayTwotransitions}, Raman spectroscopy \cite{TiSe2ChiralCDWKim2024}, second and higher harmonic generations (SHG and HHG) \cite{TiSe2nonchiralSHG} have reported evidence for additional symmetry breaking near the CDW transition temperature, beyond the standard translational symmetry breaking associated with a $2\times2\times2$ CDW formation. These findings have raised new questions regarding the nature of this unconventional CDW state \cite{chiralCDW, TiSe2unconventionCDWcal}. However, experimental results regarding the nature of the additional symmetry breaking remain controversial. For instance, while CPGE and STM suggest inversion and mirror symmetry breaking, indicative of chiral order \cite{TiSe2chiralCDWSTM, TiSe2ChiralCDWXray, TiSe2gyrotropic}, X-ray diffraction and SHG studies show that inversion symmetry is preserved \cite{TiSe2XrayTwotransitions, TiSe2nonchiralSHG}, hinting at the possible presence of ferroaxial order. Recent electron diffraction experiments also call into question previously suggested patterns for the atomic distortions responsible for the symmetry-breaking \cite{wang2025revisitingchargedensitywavesuperlattice1ttise2}. Another recent study indicates the presence of nematic fluctuations above the critical temperature via elastoresisistivity measurements \cite{lv2026nematicfluctuationmediatedsuperconductivitycuxtise2}. Moreover, several studies reveal an additional transition below the CDW transition, whose presence, symmetry, and mechanism remain elusive, further complicating the understanding of the system's symmetry-breaking behavior.
\\
\indent In the present study, we employ three techniques that each rely on measurements of physical quantities that are characterized by high-rank tensors. First, we use elastocaloric effect measurements to reveal the thermodynamic signatures of a second phase transition several Kelvin below T$_{\rm{CDW}}$. Thus, there are apparently at least two distinct phases in TiSe$_2$ which are characterized by different broken symmetries. Second, we use second harmonic generation (SHG) to definitively rule out the possibility of inversion symmetry breaking and, hence, chiral order. And third, we use elastoresistivity measurements to: (a) confirm that the broken-symmetry phase is not chiral; (b) reveal the onset of ferroaxial order; and (c) reveal how nonlinear strain can drive the motion of ferroaxial domain boundaries. In addition, above the phase transition, we also observe elastoresistivity signatures that are consistent with the presence of a diverging  susceptibility near the CDW transition temperature (T$_{\rm{CDW}}$ $\sim$ 200 K). We propose that these fluctuations also have a ferroaxial character. 
        
\section{Thermodynamic signature of an additional phase transition}
\begin{figure}[t!]
\centering
\includegraphics[trim={0cm 0cm 0cm 0cm},clip,width=0.48\textwidth]{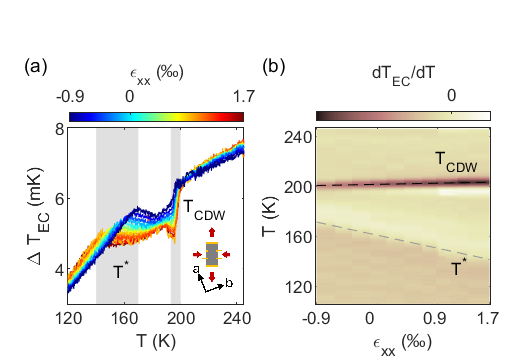}
\caption{\textbf{Thermodynamic evidence for two sequential phase transitions in 1T-TiSe$_{2}$}. (a) The elastocaloric effect as a function of temperature on a sample whose $x$-axis is oriented $22^\circ$ off the crystal's $a$-axis, measured under different DC offset strains. Uniaxial strain is applied along the $x$-axis. The shaded area indicates the transition ranges for $T_{\rm{CDW}}$ and $T^{*}$ under varied strain offsets. Inset of (a) represents the sample configuration. (b) Strain-temperature phase diagram extracted from the first derivative of the EC signal with respect to temperature. Dashed lines serve as visual guides, indicating $T_{\rm{CDW}}$ (black) and $T^{*}$ (gray) transitions.}
\label{fig:fig2}
\end{figure}


The elastocaloric effect (ECE) provides a sensitive thermodynamic probe of continuous phase transitions  \cite{ECEtransitions}. The magnitude of the ECE signal close to a continuous phase transition is proportional to the derivative of the critical temperature with respect to strain (more properly, the components of the strain tensor that are being varied in the experiment). Thus, ECE measurements are an ideal tool for detecting phase transitions that exhibit a strong strain dependence but which might otherwise have a small heat capacity anomaly relative to the total heat capacity. For example, the ECE signal in response to a strain $\epsilon_{xx}$ for temperatures close to the critical temperature $T_{c}$ is given by 

\begin{equation}
\frac{dT}{d\epsilon_{xx}} \approx \frac{C^{(c)}_{\sigma}}{C_{\sigma}}\frac{dT_{c}}{d\epsilon_{xx}},
        \end{equation}
where $\frac{dT_{c}}{d\epsilon_{xx}}$ is the strain derivative of the critical temperature,  C$^{(c)}_{\sigma}$ is the contribution of the critical fluctuations to the heat capacity, and C$_{\sigma}$ is the total heat capacity (under conditions of constant stress $\sigma$). 

Fig. \ref{fig:fig2}(a) shows representative ECE data for TiSe$_{2}$. These data were obtained by applying an in-plane uniaxial stress along a direction that is rotated 22$^{\circ}$ with respect to the crystallographic $a$ axis, thus ensuring a mixture of both $\epsilon_{x^2-y^2}$ and $\epsilon_{xy}$. Different colors represent different DC strain offsets. Phase transitions show up as sharp kinks in the data. For the largest compressive strains (blue data points) there are two clear anomalies, one just below 200 K, and one close to 175 K. For the largest tensile strains (red data points), the same two features are observed closer to 200 K and 140 K respectively. A third feature is observed for the largest tensile strains just below 200 K, but it is not clear if this is associated with a third phase transition or simply a subtle change in the shape of the anomaly close to 200 K. These features are rendered clearer by taking the first derivative of the ECE signal with respect to temperature. Plotting this derivative as a color map in the temperature-strain plane then directly yields an empirical phase diagram (see Fig. \ref{fig:fig2}(b)). The first transition, near 200 K, exhibits moderate strain dependence, while the second transition is strongly strain-tunable, with a slope of -130K/$\%$. The possible third transition for large tensile strains is evident as a white feature just below the dark feature that signals $T_\mathrm{CDW}$. Since the origin of this feature is less clear, we do not label it in the figure, focusing instead on the robust features labeled as $T_\mathrm{CDW}$ and $T^*$. Our observation of two distinct phase transitions echoes previous experimental evidence of a possible second transition below $T_{\mathrm{CDW}}$ in X-ray diffraction \cite{TiSe2XrayTwotransitions} and Raman spectroscopy\cite{TiSe2ChiralCDWKim2024}, but here revealed in a thermodynamic quantity, and also shown as a function of strain. 


\section{Evidence that the CDW phase is not chiral}
\begin{figure}[t!]
\centering
\includegraphics[trim={0cm 0cm 0cm 0cm},clip,width=0.48\textwidth]{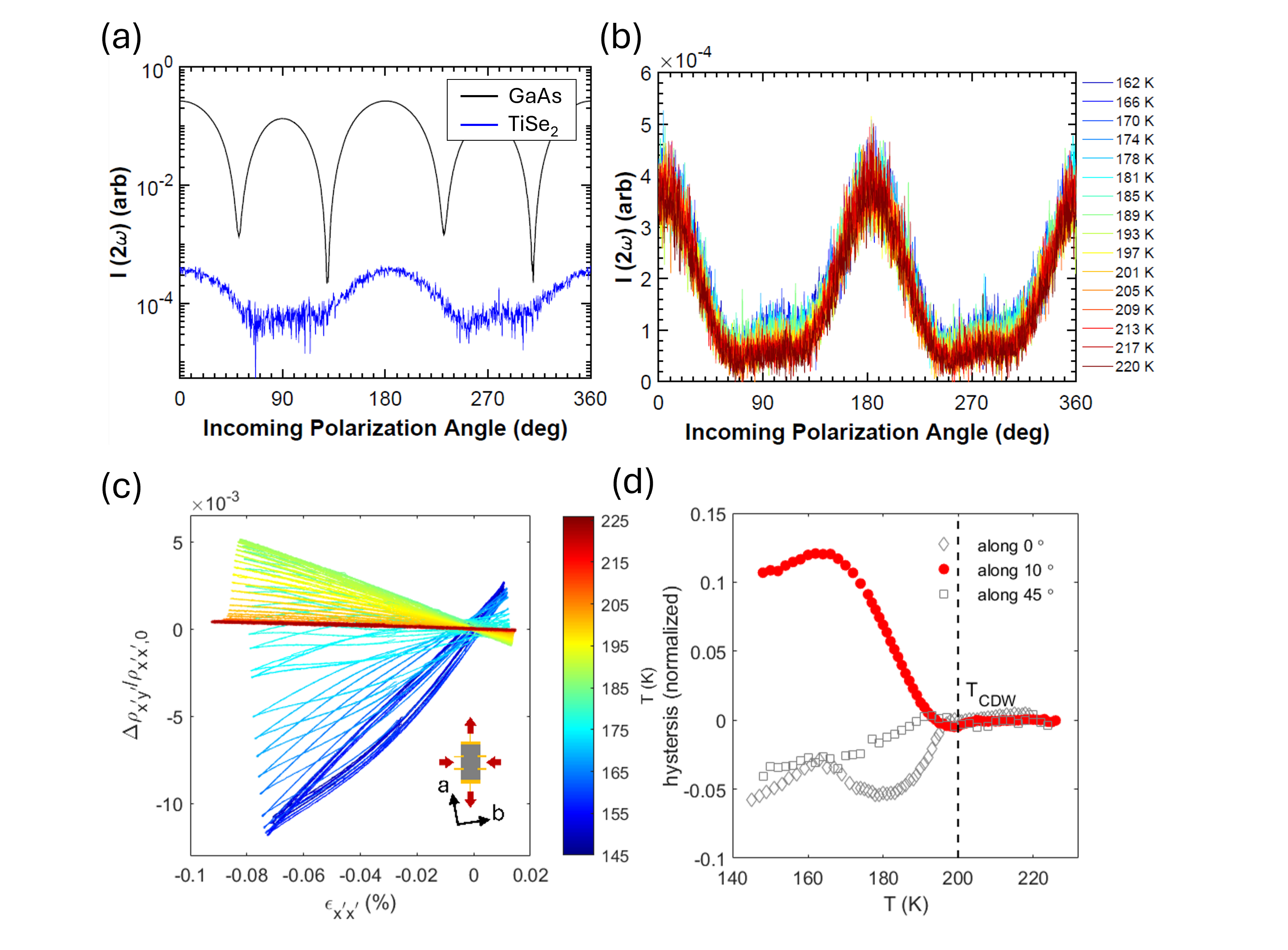}
\caption{\textbf{Evidence for non-chiral orders for both transitions in 1T-TiSe$_2$} (a) Comparison of SHG signal strengths from GaAs (black) and TiSe$_{2}$ (blue) using the same experimental parameters. The orders-of-magnitude stronger signal from GaAs indicates that the signal measured in  TiSe$_{2}$ stems exclusively from its surface. (b) Measurement of the SHG in TiSe$_{2}$ as a function of temperature shows only minimal changes, consistent with a centrosymmetric order. (c) Elastoresistivity measurements on a sample whose $x'$-axis is oriented $10^\circ$  off the crystal $a$-axis for various temperatures. Strain is applied along the $x'$-axis. Inset of (c) shows the schematics of the sample geometry and strain applied. (d) Hysteresis in $\rho_{x'y'}$  normalized by zero-strain resistivity $\rho_{0}$ as a function of temperature.}
\label{fig:fig3}
\end{figure}

To distinguish between the types of hidden order concurrent with the CDW transition -- and in particular to test whether it is a chiral order that breaks both inversion and mirror symmetries, or a ferroaxial order that breaks only vertical mirror symmetries -- we employed both high-resolution SHG measurements and domain detwinning through the application of a conjugate strain. To lowest order, SHG derives exclusively from loci of inversion symmetry breaking, including acentric bulk structures, acentric electronically ordered phases, and surfaces/interfaces~\cite{Boyd2003}. Similarly, since uniform strain is symmetric with respect to inversion it cannot couple to chiral order. Thus, experiments probing the presence or absence of domain wall motion as a consequence of uniform strain also provide a test of whether the order parameter breaks inversion symmetry or not.

Figure \ref{fig:fig3}(a) shows SHG data from TiSe$_{2}$ using a rotating incoming polarization, with the outgoing analyzer fixed in the scattering plane. This configuration provided the strongest measured response. Also shown in the same figure are data from the ``standard candle" SHG emitter GaAs using the same experimental parameters. The SHG signature from TiSe$_2$ is extremely small, and the data must be plotted with a logarithmic ordinate for both traces to be visible on the same graph. Such a small response for TiSe$_2$ can only be understood quantitatively as arising from the surface of the crystal (which naturally breaks inversion symmetry). These data provide compelling evidence that the CDW state in  TiSe$_2$ does not break inversion symmetry. Additional evidence is provided by the data in panel (b), which shows the SHG response taken as a function of temperature. The data, which span the range from 162~K to 220~K (i.e. from just above $T^{*}$ to well above $T_\mathrm{CDW}$), show only subtle changes throughout the entire range studied, and no sharp change at $T_\mathrm{CDW}$, consistent with the preservation of inversion symmetry across the first phase transitions. 

Complementary to the SHG measurements, the domain detwinning effect was studied by applying uniaxial stress oriented along various in-plane crystalline directions. We specifically studied cases in which the stress is oriented at 0$^\circ$ with respect to the crystallographic $a$-axis (i.e. only induces the strain component $\epsilon_{x^2-y^2}$); at 45$^\circ$ with respect to the $a$-axis (only strain $\epsilon_{xy}$); and at 10$^\circ$ with respect to the $a$-axis (i.e. a linear combination of $\epsilon_{x^2-y^2}$ and $\epsilon_{xy}$). As we explain in Section IV, only the cases at 10$^\circ$ and 45$^\circ$ yield a non-zero value of the effective conjugate field for the ferroaxial order, and for these two cases the sign of the effective field is reversed. In all cases, the estimated uncertainty in the alignment was $\pm 2^\circ$. We then measured the transverse elastoresistivity as a function of strain at various temperatures for each configuration. Significantly, the transverse resistivity is measured in a region of the sample for which finite element simulations indicate negligible strain gradients.  

For the elastoresistivity measurements, we use a primed coordinate system to refer to the experimental configuration, and an unprimed coordinate system to refer to the crystallographic axes. Using this convention, we apply a stress $\sigma_{x'x'}$, and measure a transverse elastoresistivity $\Delta \rho_{x'y'} / \rho_{x'x',0}$, where $\rho_{x'x',0}$ is the longitudinal resistivity for zero strain and $\Delta \rho_{x'y'}$ is measured as a function of strain $\epsilon_{x'x'}$. The strain $\epsilon_{x'x'}$ was determined using a standard capacitance technique, where we directly report the displacement measured by the calibrated capacitive sensor in the strain apparatus \footnote{Due to strain relaxation in the clamps and glue, the strain transmitted to the sample is expected to be approximately 80--90\% of the measured displacement, based on prior experience and analysis of finite element simulations.}. The zero-strain condition was estimated by comparing the temperature-dependent capacitor calibration with the known thermal expansion coefficient of the material. 

Representative data for the sample for which $\sigma_{x'x'}$ was oriented 10$^\circ$ off the crystal's $a$-axis are shown in Fig. 3(c). Strain sweeps were performed for each temperature (color scale) as the temperature was progressively reduced from 225 K to 145 K. For temperatures below $T_\mathrm{CDW}$, a pronounced hysteresis in the elastoresistivity curves is clearly observed. The magnitude of the hysteresis can be determined by integrating the area enclosed by the hysteresis loop. The temperature dependence of this quantity is shown in Fig. 3(d) for all three orientations, with the 10$^\circ$ orientation data shown in red. These data reveal a rapid increase in the hysteresis as the sample is cooled through $T_{\rm{CDW}}$. A significantly smaller hysteresis is observed for the data taken with the stress oriented at 0$^\circ$ and 45$^\circ$ with respect to the $a$-axis (gray data points).

The striking onset of hysteresis as uniform strain is swept back and forth below $T_{\rm{CDW}}$ clearly indicates that the CDW state is not chiral. Moreover, the fact that the hysteresis areas observed for strain along the $\theta=10^\circ$ and $\theta=45^\circ$ directions have opposite signs is consistent with a scenario in which the order parameter is ferroaxial. This is because, as we explain in more detail in the next section, the conjugate ferroaxial field provided by these strain directions have opposite signs. These perspectives are further emphasized by the temperature dependence of the linear and quadratic elastoresistivity coefficients, which we discuss next.

\section{Ferroxial order parameter probed by linear elastoresistivity}
\indent To further confirm the ferroaxial nature of the hidden order near T$_{\rm{CDW}}$, linear elastoresistivity was studied for samples with different orientations. To relate the former with the ferroaxial order parameter, we first perform a symmetry analysis of the resistivity anisotropy in terms of strain and ferroaxial order. For temperatures above $T_\mathrm{CDW}$, the two in-plane resistivity anisotropies, $\rho_{x^2-y^2}\equiv \rho_{xx}-\rho_{yy}$ and $\rho_{2xy}\equiv2\rho_{xy}$ transform as the two components of the $E_g$ irreducible representation (irrep) of D$_{3d}$. Similarly, the two in-plane deviatoric strain combinations, $\epsilon_{x^2-y^2}\equiv \epsilon_{xx}-\epsilon_{yy}$ and $\epsilon_{2xy}\equiv2\epsilon_{xy}$ also transform as the two components of $E_g$. Finally, the ferroaxial order parameter $\psi_{A_{2g}}$ transforms as the $A_{2g}$ irrep. Using group theory, we can express the resistivity anisotropies in powers of strain and ferroaxial order as:

\begin{align}
\rho_{x^{2}-y^{2}} & =\alpha_{1}\epsilon_{x^{2}-y^{2}}+\alpha_{2}\left(\epsilon_{x^{2}-y^{2}}^{2}-\epsilon_{2xy}^{2}\right)+\alpha_{3}\epsilon_{2xy}\psi_{A_{2g}} \nonumber \\ 
 & +\alpha_{4}\epsilon_{x^{2}-y^{2}}\psi^{2}_{A_{2g}}+2\alpha_{5}\epsilon_{2xy}\epsilon_{x^{2}-y^{2}}\psi_{A_{2g}} \nonumber \\
 &+\alpha_{6}\left(\epsilon_{x^{2}-y^{2}}^{2}-\epsilon_{2xy}^{2}\right)\psi^{2}_{A_{2g}} \label{eq:rho_1}
\end{align}

and:

\begin{align}
\rho_{2xy} & =\alpha_{1}\epsilon_{2xy}-2\alpha_{2}\epsilon_{x^{2}-y^{2}}\epsilon_{2xy}-\alpha_{3}\epsilon_{x^{2}-y^{2}}\psi_{A_{2g}} \nonumber \\
 & +\alpha_{4}\epsilon_{2xy}\psi_{A_{2g}}^{2}+\alpha_{5}\left(\epsilon_{x^{2}-y^{2}}^{2}-\epsilon_{2xy}^{2}\right)\psi_{A_{2g}} \nonumber \\
 &-2\alpha_{6}\epsilon_{x^{2}-y^{2}}\epsilon_{2xy}\psi^{2}_{A_{2g}} \label{eq:rho_2}
\end{align}where the various coefficients $\alpha_1$ through $\alpha_6$ are constants to be determined, and appear in both equations since both terms derive from the same two-dimensional irrep. 
Each term in these equations must transform in the same way, i.e., either as $E_{g}^{(1)}$ or $E_g^{(2)}$ respectively. The two significant additional pieces of group theory that are necessary to understand the origin of each one of the terms is first that for the D$_{3d}$ point group, $E_g\otimes E_g = A_{1g} \oplus \{A_{2g}\} \oplus E_g$\footnote{We use the curly brackets to indicate that the irrep $A_{2g}$ in the product decomposition corresponds to the antisymmetric combination of the two components of $E_g$}; meaning that certain quadratic combinations of strain \emph{also} transform as $E_{g}^{(1)}$ or $E_g^{(2)}$. And, secondly, the square of any fluctuating order parameter transforms as $A_{1g}$. These two constraints acting at the same time are what allows for the final terms in each of the two equations. For completeness, we note that the symmetry analysis can also be used to obtain the leading-order product of strain components that behave as a conjugate field $h_{A_{2g}}$ of the ferroaxial order parameter. We find:

\begin{equation}
h_{A_{2g}} = \epsilon_{2xy}\left(3 \epsilon_{x^2-y^2}^2 -  \epsilon_{2xy}^2 \right) \label{eq:conjugate}
\end{equation}

From Eqs. (\ref{eq:rho_1})-(\ref{eq:rho_2}), it is straightforward to obtain the off-diagonal elastoresistivity components (to linear order in strain and ferroaxial order; see also Ref. \cite{DayRoberts2025}):

\begin{align}
\left(\frac{\partial\rho_{x^{2}-y^{2}}}{\partial\epsilon_{2xy}}\right) & =\alpha_{3}\psi_{A_{2g}} - 2\alpha_{2}\epsilon_{2xy}\label{eq:elasto1} \\
\left(\frac{\partial\rho_{2xy}}{\partial\epsilon_{x^{2}-y^{2}}}\right) & =-\alpha_{3}\psi_{A_{2g}} -2\alpha_{2}\epsilon_{2xy} \label{eq:elasto2}
\end{align}

Thus, formally, in group D$_{3d}$, each of these elastoresistivity coefficients can become non-zero either because the shear strain  $\epsilon_{2xy}$ is non-zero or because the ferroaxial order parameter $\psi_{A_{2g}}$ is non-zero. In principle, these two contributions can be disentangled by considering the symmetric and antisymmetric combinations of the two elastoresistivity coefficients. In our case, we can rule out that these elastoresistivity terms are caused by unintentional shear strain present in the sample, since, as we show below, the off-diagonal elastoresistivity has a strong, order-parameter like temperature-dependence that onsets close to the CDW transition. In contrast, unintentional shear strain should give a nearly temperature-independent off-diagonal elastoresistivity. Moreover, the main contributions to the two elastoresistivity coefficients are indeed found to have opposite signs, as we show below.

Besides unintentional shear strain, the only other possible non-ferroaxial source of a non-zero off-diagonal elasto-resistivity would be from the second component of the nematic order parameter $\boldsymbol{\Phi}=(\Phi_1,\,\Phi_2)$, since $\Phi_2 \propto \epsilon_{2xy}$. This scenario can also be ruled out from our measurements. First, the CDW transition is known to preserve threefold rotational symmetry, at least close to $T_\mathrm{CDW}$. This is in agreement with recent theoretical work proposing the possible emergence of nematic order upon doping or at lower temperatures inside the CDW phase \cite{Juan2025}. Moreover, even if an undetected nematic transition were to take place, the expected non-zero component of the nematic order parameter induced would be $\Phi_1$, due to the cubic term in the Landau free energy \cite{Chakraborty2023}. 


Having established the relationship between elastoresistivity and the ferroaxial order parameter in Eqs. (\ref{eq:elasto1})-(\ref{eq:elasto2}), we now present our experimental results. Fig. \ref{fig:fig4} shows the elastoresistivity measurements on a sample oriented 45 degrees off the crystal's $a$-axis. In this configuration, the transverse resistivity $\rho_{x^{'}y^{'}}$ is proportional to the resistivity anisotropy $\rho_{xx}-\rho_{yy}$ (see Appendix I). The applied strain can be decomposed into the isotropic component, $\epsilon_{xx}+\epsilon_{yy}$, and the in-plane shear, $\epsilon_{xy}$, given the Poisson ratio of the material. In Fig. \ref{fig:fig4}(a), the elastoresistivity data are shown as a function of the anisotropic strain $\epsilon_{xy}$ at fixed temperatures (color-coded). A large nonlinear elastoresistivity response is observed near and below the CDW transition at 200K. By fitting the elastoresistivity curve with a fourth-order polynomial, one can extract the linear and quadratic elastoresistivity coefficients, which are plotted in Fig. \ref{fig:fig4}(b) and (c). The linear elastoresistivity coefficient displays a clear onset from zero below 200 K, marking the onset of ferroaxial order at (or close to) $T_\mathrm{CDW}$. We return to discuss the significance of the non-linear elastoresistivity coefficient shortly. 

An alternative way to probe the ferroaxial order parameter, according to Eq. (\ref{eq:elasto2}), is to measure the transverse resistivity induced by an in-plane anisotropic strain $\frac{\partial\rho_{xy}}{\partial\epsilon_{xx-yy}}$. Fig. \ref{fig:fig5} shows elastoresistivity measurements for a sample oriented along the crystalline $a$-axis. In this configuration, $\rho_{x^{'}y^{'}}$ directly probes the transverse resistivity $\rho_{xy}$, as a function of the in-plane anisotropic strain $\epsilon_{xx-yy}$. By fitting the elastoresistivity data in Fig. \ref{fig:fig4}(a) to a quadratic function, one can extract the linear and quadratic elastoresistivity coefficients, as shown in Fig. \ref{fig:fig5}(b) and (c). Similar to the data/configuration shown in Fig. \ref{fig:fig4}, the linear elastoresistivity shows an order-parameter-like temperature dependence across the CDW transition. The magnitude of the linear elastoresistivity coefficient $\frac{\partial\rho_{xy}}{\partial\epsilon_{xx-yy}}$ is also found to be comparable to its counterpart $\frac{\partial(\rho_{xx}-\rho_{yy})}{\partial\epsilon_{xy}}$ in Fig. \ref{fig:fig4}(b) and, more importantly, of opposite sign, as expected from Eqs.  (\ref{eq:elasto1})-(\ref{eq:elasto2}) (note the minus sign for the $y$-axes in panels (a) and (b) of Fig. \ref{fig:fig4}). However, significantly, no obvious quadratic contribution to the elastoresistivity is observed for this configuration. 

We can also use Eqs.  (\ref{eq:elasto1})-(\ref{eq:elasto2}) to elucidate the behavior observed in Fig. \ref{fig:fig3}. Applying the transformation properties of vectors and tensors, it is straightforward to derive the elastoresistivity coefficients in a coordinate system $(x',y')$  that is rotated by an angle $\theta$ with respect to the crystallographic-aligned $(x,y)$ axes. We find, for uniaxial strain applied along the $x'$ axis:

\begin{align}
\left(\frac{\partial\rho_{2x'y'}}{\partial\epsilon_{x'x'}}\right) & =-\alpha_{3}\psi_{A_{2g}}-2 \alpha_2 \epsilon_{x'x'} \sin6\theta \nonumber \\
& +2 \alpha_5 \epsilon_{x'x'} \psi_{A_{2g}}\cos6\theta \label{eq:rotated}
\end{align}
where we kept all terms linear in the ferroaxial order parameter. Similarly, in this case the conjugate ferroaxial field generated by strain is given by:

\begin{equation}
h_{A_{2g}} = \epsilon_{x'x'}^3 \sin 6\theta
\end{equation}

The expressions above show that $\left(\frac{\partial\rho_{2x'y'}}{\partial\epsilon_{x'x'}}\right)$ has a ferroaxial contribution for any rotation angle $\theta$, but a conjugate field to the ferroaxial order parameter is only present when $\theta \neq n\pi/6$. Comparing with the results of Fig. \ref{fig:fig3}(d), we conclude that the presence of hysteresis for $\theta=0^\circ$ must come from a misalignment of the axes. Moreover, in this figure, the hysteresis observed for $\theta=45^\circ$ has the opposite sign as the hysteresis observed for  $\theta=10^\circ$, in agreement with the fact that the conjugate field $h_{A_{2g}}$ has different signs in these two configurations. The data also show that the hysteresis for the  $\theta=45^\circ$ case is smaller than the hysteresis for the $\theta=10^\circ$ case, despite the conjugate field being slightly larger in the former. One possible reason for this behavior could be the non-linear term in the second line of Eq. (\ref{eq:rotated}), which vanishes for $\theta=45^\circ$ but is non-zero for $\theta=10^\circ$.

Finally, we briefly comment on the results of Fig. \ref{fig:fig4}(a)-(b) for the SHG measurements across the CDW transition. While the signal is overall small and can only be attributed to the surface, there is a small change in the magnitude of the signal as the temperature is lowered. The presence of ferroaxial order could explain this behavior. While ferroaxial order on its own does not generate a SHG signal, since it preserves inversion, it can boost the surface effect. Indeed, the surface effect can be described as an effective electric polarization $P_z$ pointing out of the plane. Using group theory, it follows that the product $\psi_{A_{2g}}P_z$ transforms as the $A_{1u}$ irreducible representation. The latter corresponds to an induced chiral order which, in turn, should change the SHG signal. The same reasoning shows why surface-sensitive probes cannot distinguish between chiral and ferroaxial order, since the explicit breaking of the horizontal mirror at the surface effectively mixes ferroaxial and chiral orders.

\section{Ferroaxial susceptibility probed by non-linear elastoresistivity}

Specific components of the rank-6 non-linear elastoresistivity tensor $\frac{d^2\rho_{ij}}{d\epsilon_{kl}\epsilon_{mn}}$ can be used to probe the ferroaxial susceptibility. Indeed, beyond ferroaxial order, Eqs. (\ref{eq:rho_1}) and (\ref{eq:elasto2}) for the resistivity anisotropies also reveal a method to probe the ferroaxial susceptibility $\langle \psi_{A_{2g}}^2 \rangle \propto \chi_{A_{2g}}$ above the CDW transition temperature. By taking the appropriate derivatives, it is evident that the two following non-linear elastoresistivity coefficients, evaluated at zero strain, can provide a window on the presence of ferroaxial fluctuations close to the ferroaxial phase transition:

\begin{equation}
    \frac{\partial^{2}(\rho_{xx}-\rho_{yy})}{\partial^{2}\epsilon_{xy}} \propto \kappa+\chi_{A_{2g}}   
\end{equation}

and 

\begin{equation}
  \frac{\partial^{2}\rho_{xy}}{\partial\epsilon_{xy}\epsilon_{xx-yy}} \propto \kappa+\chi_{A_{2g}} 
\end{equation}where $\kappa$ is some constant. It should of course be noted that in principle \emph{any} flutuating order could appear in a similar way in these expressions, since a susceptibility always transforms as the $A_{1g}$ irrep. However, sufficiently close to a phase transition that breaks only ferroaxial symmetry, the ferroaxial susceptibility must dominate over all other susceptibilities, and we therefore restrict the equations to represent only the contribution arising from the ferroaxial fluctuations.

\begin{figure}[t!]
\centering
\includegraphics[trim={0.5cm 0cm 0.5cm 0cm},clip,width=0.5\textwidth]{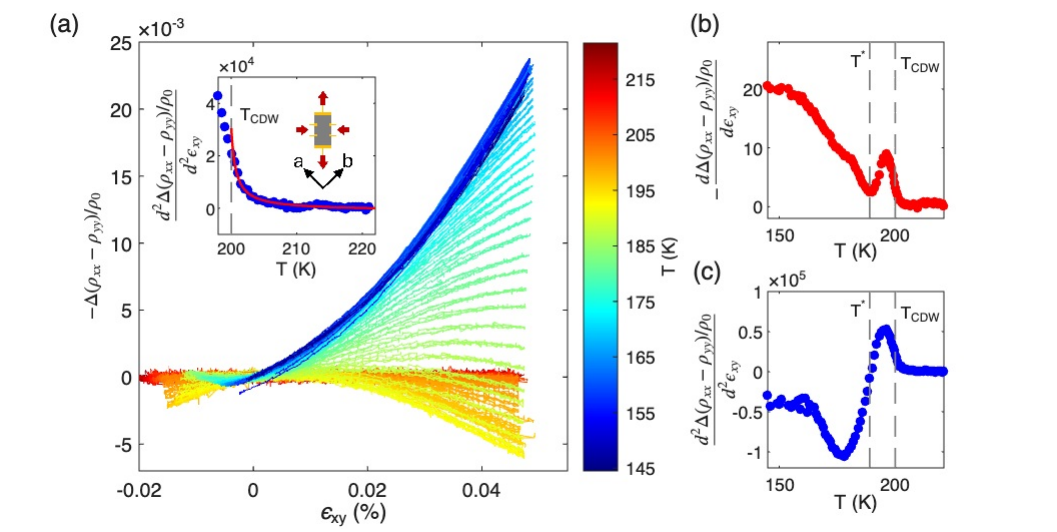}
\caption{\textbf{Evidence for ferroaxial order and a divergent ferroaxial susceptibility approaching the CDW transition in 1T-TiSe$_{2}$}. (a) Elastoresistivity measurements on a sample whose $x'$-axis is oriented $45^\circ$ off the crystal $a$-axis for various temperatures. (b) Linear and (c) quadratic elastoresistivity coefficient as a function of temperature, extracted by fitting the elastoresistivity data in panel (a). Inset of (a) shows the quadratic elastoresistivity coefficient as a function of temperature above $T_{\rm{CDW}}$ and the schematics of the sample geometry. The red curve represents a fit to the Curie–Weiss equation, yielding a Weiss temperature of 199.3 $\pm$ 0.2 K, very close to $T_{CDW}$. Note the negative sign for the y-axis labels in panels (a) and (b).}
\label{fig:fig4}
\end{figure}

\begin{figure}[t!]
\centering
\includegraphics[trim={0.5 0cm 0.5cm 0cm},clip,width=0.5\textwidth]{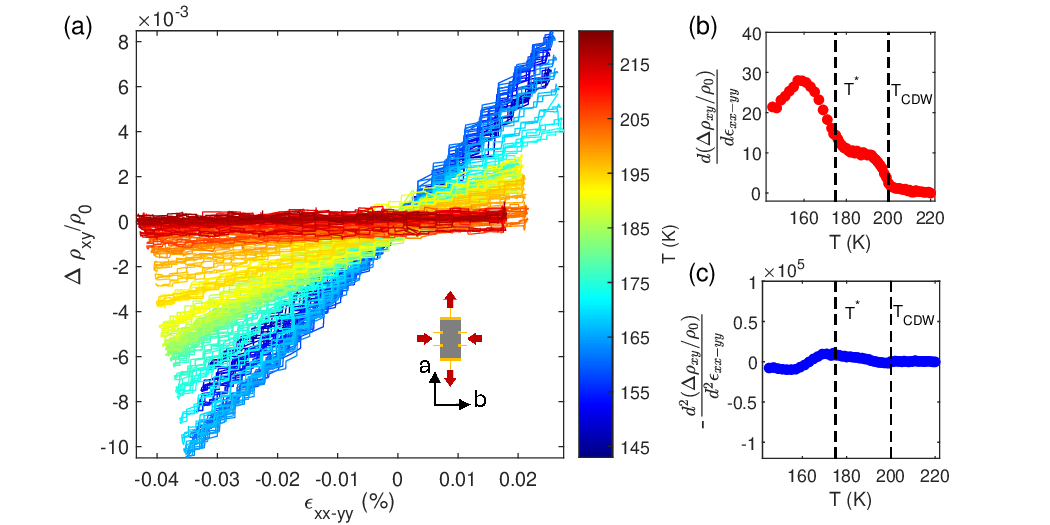}
\caption{\textbf{Evidence for a ferroaxial order parameter across the CDW transition in 1T-TiSe$_{2}$} (a) Elastoresistivity measurements on a sample whose $x'$-axis is oriented parallel to the crystal axis $a$ for various temperatures. Inset of (a) shows the schematics of the sample geometry and strain applied. (b) Linear and (c) quadratic elastoresistivity coefficients $\frac{\partial(\Delta \rho_{xy}/\rho_{0})}{\partial\epsilon_{xx-yy}}$ and $\frac{\partial^{2}(\Delta \rho_{xy}/\rho_{0})}{\partial\epsilon_{xx-yy}^{2}}$, respectively, as a function of temperature and extracted by fitting the elastoresistivity data in panel (a).}
\label{fig:fig5}
\end{figure}

Motivated by these considerations, we return to consider the non-linear contributions to the elastoresistivity shown in Figures \ref{fig:fig4} and \ref{fig:fig5}. First, considering the data shown in Fig.  \ref{fig:fig4}, for which  $(\rho_{xx}-\rho_{yy})$ is measured as a function of $\epsilon_{xy}$, the quadratic component of the elastoresistivity $\frac{\partial^{2}(\rho_{xx}-\rho_{yy})}{\partial^{2}\epsilon_{xy}}$ is found to grow rapidly upon approaching $T_\mathrm{CDW}$ from above (see inset to Fig.  \ref{fig:fig4}(a)). Indeed, the data can be fit over a 20 K range to a Curie-Weiss functional form (red line in the inset to Fig  \ref{fig:fig4}(a)), though this is only suggestive of a divergence given the relatively small range of reduced temperatures that these data span.  

In contrast, considering the data shown in Fig.  \ref{fig:fig5}(a), for which $\rho_{xy}$ is measured as a function of $\epsilon_{xx-yy}$, no obvious quadratic elastoresistivity response was observed for this configuration. This is consistent with the expectations from Eq. (\ref{eq:elasto2})  if the dominant fluctuations have a ferroaxial character, since a combination of $\epsilon_{xx-yy}$ and  $\epsilon_{xy}$ is needed to probe the ferroaxial susceptibility in the transverse resistivity $\rho_{xy}$. Small misalignment of the crystal axes and/or the presence of any unintentional shear strain are presumably then responsible for any small induced anisotropy observed in this regime. 
Thus, while not a smoking gun on its own, the non-linear elastoressistivity provides additional compelling evidence, when combined with the linear elastoresistivity measurements, for the presence of ferroaxial fluctuations above $T_\mathrm{CDW}$.

\section{Discussion}
Using elastocaloric measurements as a sensitive thermodynamic probe, we have revealed the presence of at least two distinct phase transitions in 1T-TiSe$_2$, one at $T_{\rm{CDW}} \sim 200$ K, and a second strain-dependent transition $T^{*}$ whose range spans from 140 K to 190 K for the range of strains investigated. Our study has concentrated on establishing the symmetry of the higher temperature phase that onsets at $T_{\rm{CDW}}$. Our SHG measurements rule out the possibility of chiral order, an erstwhile candidate suggested by previous surface-sensitive probes \cite{TiSe2gyrotropic, TiSe2chiralCDWSTM}. Meanwhile, the striking observation of an anomalous antisymmetric off-diagonal response associated with the $x^2-y^2$ and $xy$ components of the resistivity and strain tensors in this phase provides compelling evidence for a ferroaxial component to the CDW order. Additional supporting evidence is found in the observation of hysteresis in response to sweeping back and forth of the conjugate effective field in the ordered state, and the observation of a Curie-Weiss-like divergence of the orthogonal non-linear elastoresistivity upon approaching the phase transition from above. 
\\
\indent Although extensive reports have suggested mirror symmetry breaking in this material, to complete our discussion, we briefly consider another possible symmetry-breaking phase -- the 3-state Potts nematic order, which breaks the three-fold rotational symmetry rather than the mirror symmetry \cite{Chakraborty2023}. Although the difference between these two symmetry-breaking phases in the elastoresistivity response of this material is subtle, we identify two key pieces of evidence that suggest it is unlikely to be a nematic order.
\\
\indent The first piece of evidence is the absence of nematic susceptibility in elastocaloric measurements. Such a divergent susceptibility has been observed in other systems hosting 3-state Potts nematicity \cite{FPSe3PottsNematicity}. However, in our elastocaloric measurements (Fig. \ref{fig:fig2}(a)), no feature typically associated with a susceptibility is observed above $T_{\rm{CDW}}$. This absence of susceptibility in elastocaloric measurements is consistent with ferroaxial order, where a higher order conjugate strain, see Eq. (\ref{eq:conjugate}), is needed to couple to its order parameter, leading to an infinitesimally small effect in ECE. The second piece of evidence comes from the orientation dependence of the hysteresis size. For a 3-state nematic order, hysteresis should be most pronounced when stress is applied along the crystal axis, as it would favor one of the three nematic domains. In contrast, for a ferroaxial order, the conjugate strain is zero when the sample is aligned with the crystal axis, leading to a diminished hysteresis in samples oriented along the $a$-axis compared to other configurations.
\\
\indent While the symmetry of the hidden order associated with the higher temperature CDW phase has been identified, the origin of this ferroaxial order remains unclear. Potential explanations, such as the role of relative phases of atomic displacements and/or other orbital textures within the CDW state, have been proposed \cite{wang2025revisitingchargedensitywavesuperlattice1ttise2}. Intriguingly, evidence for similar ferroaxial CDW states has also been observed in other material systems beyond 1T-TiSe$_{2}$, including most recently $R$Te$_{3}$ (where $R$ is a rare earth ion) \cite{Ken2025,Rafael2024PRB}. The discovery of the ferroaxial CDW state in TiSe$_{2}$ thus provides a valuable platform for investigating its microscopic mechanism, which could shed light on analogous unconventional CDW states in other systems.
\\
\indent An intriguing open question remains regarding the nature of the second transition, at $T^*$. Our SHG data show no signature of inversion symmetry breaking down to 162~K, which rules out chiral order. Another possibility is a 3-state Potts nematic order, suggested by the strong strain dependence and resistivity anisotropy associated with the second transition. This scenario would be consistent with the observation of a splitting of the $E_{g}$ mode in Raman spectroscopy \cite{TiSe2ChiralCDWKim2024} and an anisotropic component in HHG near 160K \cite{TiSe2HHGnonchiral}. It is also important to consider that the second transition may not be continuous; if it is a first-order transition, this opens up a broader range of possibilities, unconstrained by the space group of the higher temperature CDW phase.
\\
\emph{Note added}: A similar experimental investigation of the elastoresistivity of 1T-TiSe$_2$ has been carried out concomitantly to this work in Ref. \cite{jiunhawFAarxiv}. The authors of that study, which include one of the co-authors of the present paper (R.M.F.), draw very similar conclusions to our own as to the presence of ferroaxial order in TiSe$_2$.
\\

        
\section{Experimental methods}
Single crystals of 1T-TiSe$_{2}$ were synthesized by the chemical vapor transport method with iodine as the transport agent. Stoichiometric amounts of titanium sponge (99.95\%) and selenium powder (99.999\%) were mixed with iodine (1$\sim$mg/cm$^3$) and sealed in quartz tubes ($\sim$ 8 inches in length) under high vacuum. The tubes were then placed in a three-zone horizontal tube furnace with the charge positioned near one of the furnace thermometers. Sizeable crystals ($10\times10\times0.3~\text{mm}^3$) were obtained after gradually heating the precursor to 650$^\circ$C at the source end and 575$^\circ$C at the sink end, dwelling for two weeks, and cooling down to room temperature. The resulting single crystals of TiSe$_{2}$ were aligned using an Xpert2 single crystal X-ray diffractometer by identifying the [101] Brag peak. 

Elastoresistivity measurements were performed by cutting the pre-aligned samples in a Hall-bar geometry along the desired direction, attaching it to the Razorbill CS100 strain cell, and measuring its transport properties, $\rho_{x'x'}$ and $\rho_{x'y'}$, using a Physical Properties Measurement System (PPMS). The measured transport properties $\rho_{x'x'}$ and $\rho_{x'y'}$ in the experimental coordinate system can be converted to the resistivity anisotropic $\rho_{xx}-\rho_{yy}$ and transverse resistivity $\rho_{xy}$ in the crystal coordinate system, provided the sample direction. The measurement's strain zero point is determined by matching the sample length indicated by the capacitance sensor with the zero strain value calculated based on the apparatus's zero-strain gap size and the material's thermal expansion coefficient. 

Second harmonic generation measurements were taken at $45^\circ$ angle of incidence using incoming wavelength $\lambda = 1300$~nm and 1.03~mW of incident power. The data were acquired as a function of incoming polarization angle using four different polarization combinations, specifically: incoming polarization rotating, outgoing analyzer rotating parallel to the incoming polarization ($I_\parallel$); incoming polarization rotating, outgoing analyzer rotating perpendicular to the incoming polarization ($I_\perp$); incoming polarization rotating, outgoing analyzer fixed in the scattering plane ($I_H$); and incoming polarization rotating, outgoing analyzer fixed perpendicular to the scattering plane ($I_V$).

\section{Acknowledgments}
The authors thank J.-H Chu for sharing a copy of their manuscript prior to jointly submitting our two papers. We also thank M. Gastiasoro, F. de Juan, and I. Maccari for discussions. Elastoresistance and elastocaloric effect measurements at Stanford were supported by the Department of Energy, Office of Basic Energy Sciences, under Contract No. DE-AC02-76SF00515. QJ was partially supported by the Gordon and Betty Moore Foundation Emergent Phenomena in Quantum Systems Initiative through Grant GBMF9068. Second harmonic generation measurements at Temple University (A. D and D. H. T.) were funded  by the National Science Foundation under Award No. DMR-1945222. The work at the University of Minnesota (E. D. R. and T. B.) was supported by the National Science Foundation through the University of Minnesota MRSEC under Award No. DMR-2011401. EDR also acknowledges partial support from NSF Award No. DMR-2206987.


\section{Appendix: Transformation of the strain and resistivity tensors under coordinate system rotation}
Strain $\epsilon_{ij}$ and resistivity $\rho_{kl}$ are both rank-two tensors. The in-plane components can be expressed  
\begin{equation}
    \epsilon=\begin{pmatrix} 
    \epsilon_{xx} & \epsilon_{xy}\\
    \epsilon_{yx} & \epsilon_{yy}
    \end{pmatrix}
    \end{equation}

and

    \begin{equation}
    \rho=\begin{pmatrix} 
    \rho_{xx} & \rho_{xy}\\
    \rho_{yx} & \rho_{yy}
    \end{pmatrix}
\end{equation}
where $x$ and $y$ coordinate axes are orthogonal and are defined relative to  given crystallographic directions. We specifically choose the $x$ direction to lie along the crystallographic $a$-axis. To probe hidden orders with unknown symmetry breaking, we initially assume no crystal symmetries that would otherwise constrain the terms in the resistivity tensor. 

For the experiments, both uniaxial stress and current are applied and measured at an angle $\theta$ with respect to the crystalline $a$-axis (note that current and stress are applied in exactly the same direction in this experiment by construction). We use primed indices to denote values of the components of the strain and resistivity tensors in this rotated coordinate system. Figures \ref{fig:fig4} and \ref{fig:fig5} report values of the resistivity and strain in the frame of the crystal (unprimed coordinates). For Fig \ref{fig:fig5} this does not require any transformation; the measurement is performed in the unprimed coordinate system. However, for Fig.  \ref{fig:fig4}, this requires transforming the strain and resistivity tensors between these two coordinate systems.  

        

Let $\left(x',y'\right)$ be the coordinate system rotated with respect to $\left(x,y\right)$ by an arbitrary in-plane angle $\theta$ with respect to the $x$ axis. Then, defining the rotation matrix (we are rotating the coordinate axes around the $z$-axis):

\begin{equation}
R_{\theta}=\left(\begin{array}{cc}
\cos\theta & \sin\theta\\
-\sin\theta & \cos\theta
\end{array}\right)
\end{equation}
we have $\mathbf{r}'=R_{\theta}\mathbf{r}$, $\hat{\epsilon}'=R_{\theta}\hat{\epsilon}R_{\theta}^{T}$, and $\hat{\rho}'=R_{\theta}\hat{\rho}R_{\theta}^{T}$. Explicitly:

\begin{widetext}

\begin{equation}
\left(\begin{array}{cc}
\epsilon_{xx} & \epsilon_{xy}\\
\epsilon_{xy} & \epsilon_{yy}
\end{array}\right)=\left(\begin{array}{cc}
\epsilon_{x'x'}\cos^{2}\theta+\epsilon_{y'y'}\sin^{2}\theta-\epsilon_{x'y'}\sin2\theta & \epsilon_{x'y'}\cos2\theta+\frac{1}{2}\left(\epsilon_{x'x'}-\epsilon_{y'y'}\right)\sin2\theta\\
\epsilon_{x'y'}\cos2\theta+\frac{1}{2}\left(\epsilon_{x'x'}-\epsilon_{y'y'}\right)\sin2\theta & \epsilon_{x'x'}\sin^{2}\theta+\epsilon_{y'y'}\cos^{2}\theta+\epsilon_{x'y'}\sin2\theta
\end{array}\right)
\end{equation}
and:

\begin{equation}
\left(\begin{array}{cc}
\rho_{xx} & \rho_{xy}\\
\rho_{xy} & \rho_{yy}
\end{array}\right)=\left(\begin{array}{cc}
\rho_{x'x'}\cos^{2}\theta+\rho_{y'y'}\sin^{2}\theta-\rho_{x'y'}\sin2\theta & \rho_{x'y'}\cos2\theta+\frac{1}{2}\left(\rho_{x'x'}-\rho_{y'y'}\right)\sin2\theta\\
\rho_{x'y'}\cos2\theta+\frac{1}{2}\left(\rho_{x'x'}-\rho_{y'y'}\right)\sin2\theta & \rho_{x'x'}\sin^{2}\theta+\rho_{y'y'}\cos^{2}\theta+\rho_{x'y'}\sin2\theta
\end{array}\right)
\end{equation}


\end{widetext}

In the experiment, uniaxial stress $\sigma_{x'x'}$ induces in-plane strains $\epsilon_{x'x'}$ and $\epsilon_{y'y'}$. For an arbitrary coordinate system rotation, $\theta$, the rank 4 elastic stiffness tensor $C_{ijkl}$ transforms to $C_{i'j'k'l'}$. For rotations away from principal axes, uniaxial stress can also induce shear in the primed coordinate system. However, shear strain $\epsilon_{x'y'}$ is heavily suppressed due to the high stiffness of the strain cell used in the experiment, and also by compatibility relations of the material (gradients in shear strain necessarily induce significant tensile and compressive strains, which are very energetically costly). We can therefore neglect any shear in the experimental frame of reference (i.e. we assume $\epsilon_{x'y'} = 0$). The longitudinal strain $\epsilon_{x'x'}$ is measured (deduced from displacement of the jaws of the strain cell), while $\epsilon_{y'y'}$ is deduced from $\epsilon_{x'x'}$ based on the known Poisson ratio of the material. Finally, the transverse resistivity $\rho_{x'y'}$ is measured. All of the measured/deduced quantities are then transformed back into the unprimed coordinate system using the formulae above:
\begin{align}
\epsilon_{x^{2}-y^{2}} & =(\epsilon_{x'x'}-\epsilon_{y'y'})\cos2\theta\nonumber \\
\epsilon_{2xy} & =(\epsilon_{x'x'}-\epsilon_{y'y'})\sin2\theta\nonumber \\
\rho_{x^{2}-y^{2}} & =(\rho_{x'x'}-\rho_{y'y'})\cos2\theta\nonumber - 2\rho_{x'y'}\sin2\theta\nonumber \\
\end{align}

For the experiment shown in Fig \ref{fig:fig4}, $\theta = 45^o$, and $\rho_{x^{2}-y^{2}}=-2\rho_{x'y'}$. \\

\bibliography{TiSe2FerroaxialOrder.bib}

\end{document}